Michele Tucci

DEPARTMENT OF PUBLIC ECONOMICS
FACULTY OF ECONOMICS
UNIVERSITY OF ROME "LA SAPIENZA"

# EVOLUTIONARY SOCIOECONOMICS: A SCHUMPETERIAN COMPUTER SIMULATION

*Economic Wars in the Cyberspace*

# Abstract

The following note contains a computer simulation concerning the struggle between two companies: the first one is "the biggest zaibatsu of all", while the second one is "small, fast, ruthless". The model is based on a neo-Schumpeterian framework operating in a Darwinian evolutionary environment. After running the program a large number of times, two characteristics stand out:

- *There is always a winner which takes it all, while the loser disappears.*
- *The key to success is the ability to employ efficiently the technological innovations.*

The topic of the present paper is strictly related with the content of the following notes:

Michele Tucci, *Evolution and Gravitation: a Computer Simulation of a Non-Walrasian Equilibrium Model*.

Michele Tucci, *Oligopolistic Competition in an Evolutionary Environment: a Computer Simulation*.

The texts can be downloaded respectively at the following addresses:
http://arxiv.org/abs/cs.CY/0209017
http://arxiv.org/abs/cs.CY/0501037
These references include some preliminary considerations regarding the comparison between the evolutionary and the gravitational paradigms and the evaluation of approaches belonging to rival schools of economic thought.

**Information on the author**: MICHELE TUCCI, Department of Public Economics, Faculty of Economics, University of Rome "La Sapienza", Italy.
Email: tucci@dep.eco.uniroma1.it
Web page: http://dep.eco.uniroma1.it/~tucci/
Or: http://xoomer.virgilio.it/michele_tucci/

# *EVOLUTIONARY SOCIOECONOMICS: A SCHUMPETERIAN COMPUTER SIMULATION*

MICHELE TUCCI

## 1. Introduction

William Gibson, in his short story "New Rose Hotel", depicts the war for technological supremacy between two transnational companies. The first one, the Japanese Hosaka, is described as "the biggest zaibatsu of all", while about its Germanic competitor he specifies that "Maas was small, fast, ruthless. An atavism. Maas was all Edge". The story goes on narrating the fierce competition between the two conglomerates, conducted mainly by abducting – and/or terminating – the best scientific researchers available on the market. Of course that's fiction! Reality is much worse… What Gibson doesn't tell us is which one is going to be the final winner, the only one left. The struggle goes on and on…
From the point of view of the economist, Gibson's novel brings in mind a traditional fragment of economic theory: Schumpeter's treatment of the evolution of the firm productive technologies by means of the struggle between a defender and a challenger. It should be noted that Schumpeter's approach is clearly based on the Darwinian evolutionary model: the process is centered on the binomial "mutation – selection". The first phase – the mutation – includes the appearance of new technologies which have been tested in laboratories and perhaps in experimental plants, but have never been used in standard production lines. In order to allow the challenging firm to produce goods with a new technology, a merchant bank must be willing to run the risk to finance the innovation. If this happens, the challenger is ready to start the fight and the second phase – the selection – will start. The traditional criterion to pick the winner is based on production cost: if the new technology is associated to a lower level of such a parameter, then the challenger should win. Vice versa, the defendant will be able to keep its grip on the market. In the real world the situation is rather different: lower cost may be a necessary condition, but not a sufficient one. The game that is played by the two competitors is critical: each move can advantage one or the other up to the point that sometimes a superior technology may be not enough to win, if the management of the challenging firm is very poor. Examples of such a type of events are not rare. As it is for the appearance of a mutation, selection is a rather chaotic process and it's not easy to model it. To attempt such a task, the Darwinian evolutionary approach must be employed. A computer simulation which is based on the de Finetti – Simon principia – and it's carried on by skilled economists and computer experts – holds a chance to reach a certain degree of forecasting reliability, while the same cannot be affirmed if we employ the traditional gravitational approach. While the limits of the gravitational approaches are evident, there is no doubt that the model which will eventually lead us to a scientific understanding of the economic and social phenomena is the Darwinian evolutionary framework.

Still, we are left with our curiosity: which one of the two cited firms is going to win? Of course, it depends on the contest: we have to employ a vast set of variables in order to outline the environment where the game is going to be played and we have to provide a reconstruction of the behavior patterns of each company. Then we can let the simulation run and see...

That's what will be carried out in the next paragraph. A very simplified simulation – at the level of a "proof of concept" – has been coded by the author in C++ using MS Visual Studio 6.0. The nature of each company – the defender Hosaka (H) and the challenger Maas (M) – has been sketched by defining a limited number of parameters and behavioral functions. H is big – and therefore at the beginning of each simulation cycle its market share will be set equal to two thirds of the total amount – but its size implies a degree of lethargy – i.e. it's slow in exploiting the advantages of technological progress. On the other side, M has the talent to ride the waves of innovation – thus the parameter defining such feature will be set for M to a higher value than for H. Moreover, M is aggressive and this trait will be embodied in two especially designed functions: "Protect your success" and "Attack when the adversary hesitates". After running the program a large number of times, two features stand out:

- *There is always a winner which takes it all, while the loser disappears.*
- *The key to success is the ability to employ efficiently the technological innovations.*

Moreover, the simulation shows some interesting stability properties. In the following paragraph we'll go a bit more deeply into the matter…

## 2. The Simulation

Let's examine the details of the simulation. In compliance with Simon's principles, the time cycle of the simulation will be structured in a finite sequence of Hicksian "weeks" – generally a set of thirty time periods. At times, longer cycles have been employed, up to one hundred periods. The agents that will appear in the model are the following.

*The technical progress (TP)*. This agent will represent the complex process of providing innovative technologies which bear the potentiality to beat the actual markets – i.e. supplying new commodities and/or new productive methods which can overcome the present ones in the preferences of the consumers and/or by lowering the production costs. Of course, in order to turn a potential advantage into an actual one, an investment is needed. Thus, TP is useful if it can be linked to a given amount of money. As it will be shown soon, a specific agent, the banker, will take care of this problem. Since building a model of the inner structure of TP is beyond the scope of the present simulation, at the beginning of each period, TP will be represented by a random number from 1.0 to 10.0 obtained by the usage of the standard C++ function *rand()*.

*The defender company (H)*. Since H is "the biggest zaibatsu of all", at the beginning of the simulation we will set its market share equal to 0.75, while the challenger's one will amount to 0.25. During each period, the level of the H profit will be set to depend on two variables: the value of TP and the amount of investment. The mathematical expression will be shaped as the following:

(I)     (H profit) = (TP)$^{(H\ exp)}$ * (H investment)

*The challenger company (M)*. Starting with a minority quote of the market share, in order to try and win M must behave aggressively. But we already know that "Maas was small, fast, ruthless. An atavism. Maas was all Edge". Therefore, such behavior is exactly what M is due to embody. Let's model such a feature of M. Firstly, during each period the level of M profit will be defined by an expression similar to (I):

(II)    (M profit) = (TP)$^{(M\ exp)}$ * (M investment)

When running the simulation program, we will set (M exp) > (H exp), thus modeling the fact that since M is smaller than H and a lot more agile, it will be able to use technical progress in a more profitable way. Moreover, two more functions will be implemented: "Protect your success" and "Attack when the adversary hesitates", both stressing the aggressive nature of M. The first function will set the following rule: "if for two consecutive periods M market share has been increasing, then in the next period an extra amount of money should be invested to keep the lead". The second function can be summarized in the following way: "If in the previous period the H market share has been decreasing, then in the next period an extra amount of money should be invested in order to take advantage of the adversary's weakness". It should be noted that since we are not going to model the financial behavior of the two companies in details, it will be supposed that the above mentioned extra money will be supplied by an unspecified source within M – for example by private investors who believe that M will be the winner.

*The banker (B)*: the money provider. The banker is the "deus ex machina" in the Schumpeterian reconstruction of the evolution of the productive structures. Money is necessary to turn a potentially winning new technology into standard production lines that can provide commodities to be sold on the market. In fact, the evolution of the productive apparatus always implies the presence of those two factors: promising inventions and discoveries coupled with capital to put them to work. While the availability of the first element is determined by a complex process involving rather intangible elements, the second one, capital, is controlled by the obvious economic logic: expected profit. Thus, in our basic simulation we will suppose that the banker should behave in a conservative way, by lending money in proportion with the level of profit which each firm earned during the previous period of time. The rationale is that in general terms more profit means more possibilities to reimburse the loans and therefore more possibilities for the banker to obtain the expected financial profit. The mathematical expressions are the following, holding for both firms:

(III)
(investment) = (money from the B)

(money from the B) = (constant) * (profit of the previous period)$^{(money\ exp)}$

*The market share (MS)*: who is winning and who is losing. In the real world the market share of a company will depend on objective elements – such as the joint competitive advantage of its productive and financial structures on one side and of its products on the other side – and on subjective ones, such as the strategies carried on by the management. Since in our simulation the scope of both firms is global domination – i.e. kicking the adversary out of existence – we will suppose that every available asset of the company – i.e. the total amount of the profit for H, the same plus the eventual outcome of the two functions "Protect your success" and "Attack when the adversary hesitates" for M – will be used to increase its own market share. For example, this can be done by financing a suitable advertising campaign and/or by lowering the actual selling price of the produced commodity. The expressions are the following:

(IV)   (H MS) = (constant) * (H profit)

(V)    (M MS) = (constant) * [(M profit) + ("Protect your success") + ("Attack when the adversary hesitates")]

By setting:

(VI)   (constant) = 1 / [(H MS) + (M MS)]

the relative quotas of the firm market shares will be calculated.

The structure of the simulation is represented in diagram 1.

**1 – The structure of the simulation**

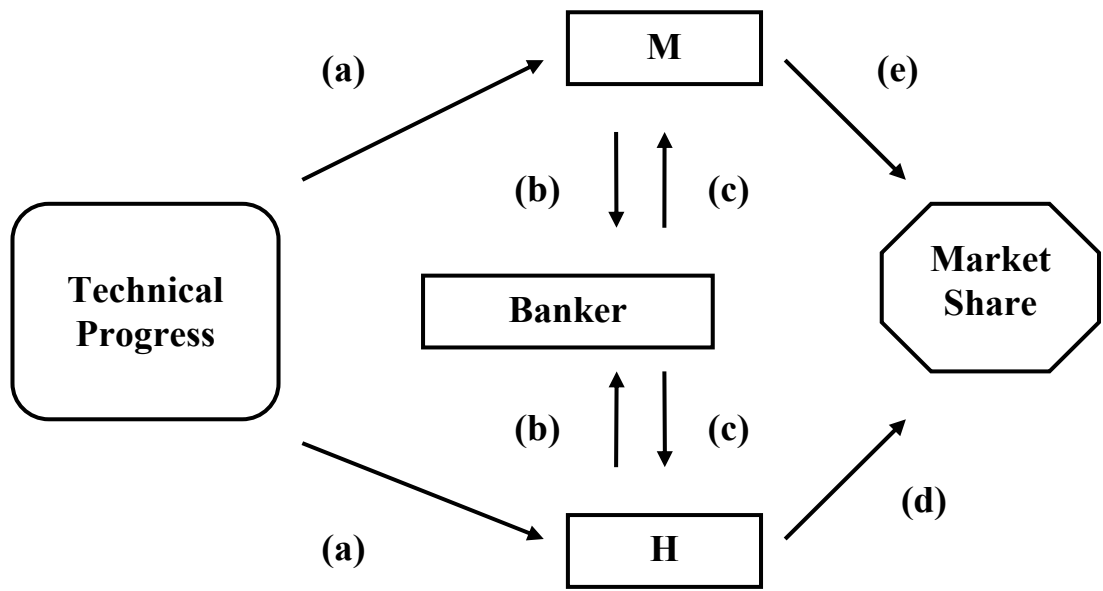

The arrows in diagram 1 indicate the direction of the links among the agents. Let's see the details.

(start). At the beginning of each time period TP defines the actual level of the technological progress by using of the standard C++ function *rand()*.

(a). TP communicates to H and M the actual level of technological progress.

(b). H and M inform B about their profit levels relative to the previous time period.

(c). B lends money to H and M in the quantity specified by expression (III), thus setting the investment level. Therefore, by using respectively expression (I) and (II), H and M are able to determine its own level of profit for the actual period of time.

(d). MS sets H market share according to expression (IV).

(e). MS sets M market share according to expression (V).

(end). MS define the relative quotas of the market shares by using normalization (VI).

Then the iteration moves to the next time period and the above sequence of actions is repeated. The program terminates at the last period of the time cycle.
The simulation has been run for a large set of values for the parameters. The time cycle of the model varied from thirty periods up to one hundred. At times, in order to test a specific pattern, instead of using random values TP has been exogenously fed for every time period. At the end of such a thorough testing, a remarkable feature related to structural stability stands out:
- ***In every case there is a clear winner who takes it all, while the loser's market share goes to zero.***

Moreover, if we examine the graphs of the market shares of the two companies through time, we notice that they assume only two shapes. An example of such a feature is shown in graphs 2 and 3: the first one refers to the case when H wins, while in the second one M does. Graphs 4 and 5 represent the values of TP relative to the contests illustrated respectively in graph 2 and 3. Such an outcome ought to be considered as the optimal one, since it can be concluded that the simulation represents clearly and without any ambiguity the fragment of reality under examination: a merciless struggle between a defendant and a challenger. Among the parameters, a critical role is played by the exponents which appear in expression (I) and (II). They express the ability of each firm to usefully employ the new technologies available during the current time period. As it has already been pointed out, the exponent M exp, relative to the firm M, should be greater than H exp, which pertains to H, since M is "small, fast, ruthless" while H is "the biggest zaibatsu of all".
- ***Let's consider the set of all the parameters of the simulation, with the exception of H exp and M exp. For all the tested values of the parameters included in such a set, there is always a couple of values for H exp and M exp, with M exp > H exp, for which if we run the program enough times victories and defeats are evenly split between the two companies. Moreover, if we leave the remaining parameters unchanged and increase M exp, them M tends to win always, and vice versa.***

This property conveys the relevance of the simulation: the ability of using efficiently the resources that technical progress provides is the key to dominate the markets. Therefore our simulation can be considered a faithfully expression of Schumpeter's vision. It's a new tool ready to be developed from the stage of a "proof of concept" to a fully effective operative utensil, useful in tackling with the economic phenomena of the real world.

It should be noted that the example illustrated in graphs 2 - 5 has been obtained from a set of values for the parameters such that, after running the simulation enough times, victories and defeats are evenly split between the two firms. Moreover, if we give a look to graphs 4 and 5, which show the values of TP in the case of the victory respectively of H and M, at first sight we are unable to deduct the reason why in the first context H should win, while in the second one M does. Definitely, the shapes of the graphs are devoid of any meaningful suggestion. Thus, even at the elementary level of a "proof of concept" the simulation holds the power to create dynamical trends which cannot be deducted from a priori considerations. The chaotic interaction between just a few basic agents is able to shed light on critical phenomena of the economic world that are difficult to tackle with the traditional formulas. What will eventually happen if we move to a realistic full scale simulation?

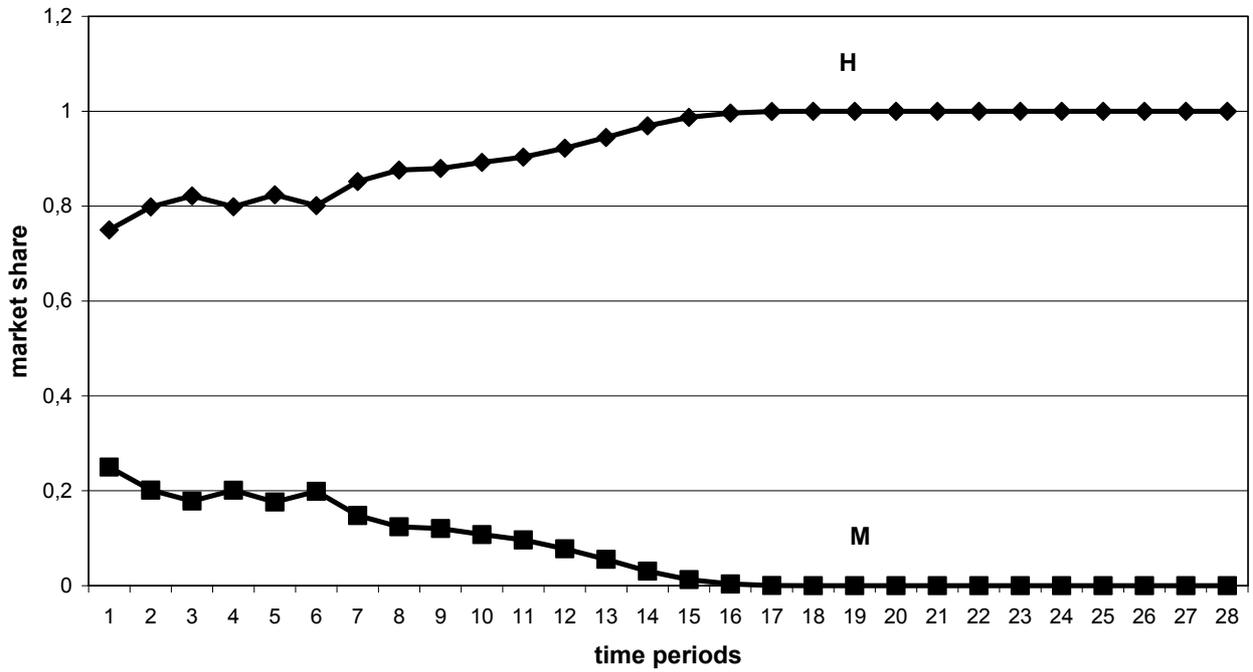

2 - H wins!

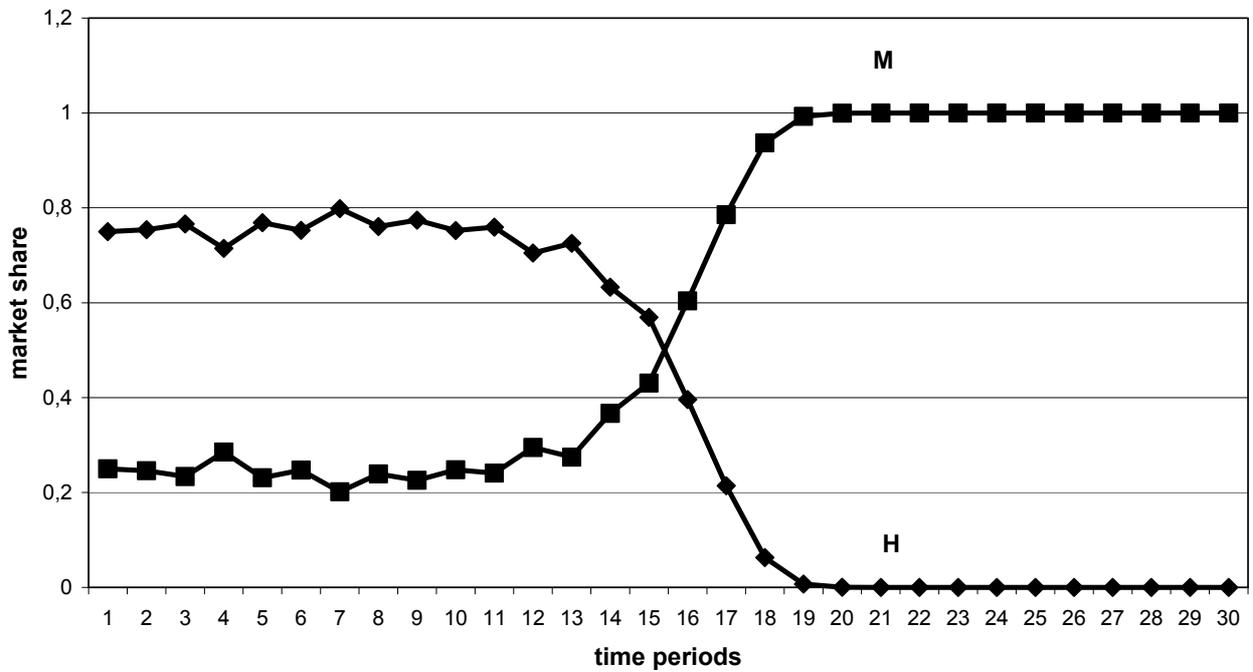

3 - M wins!

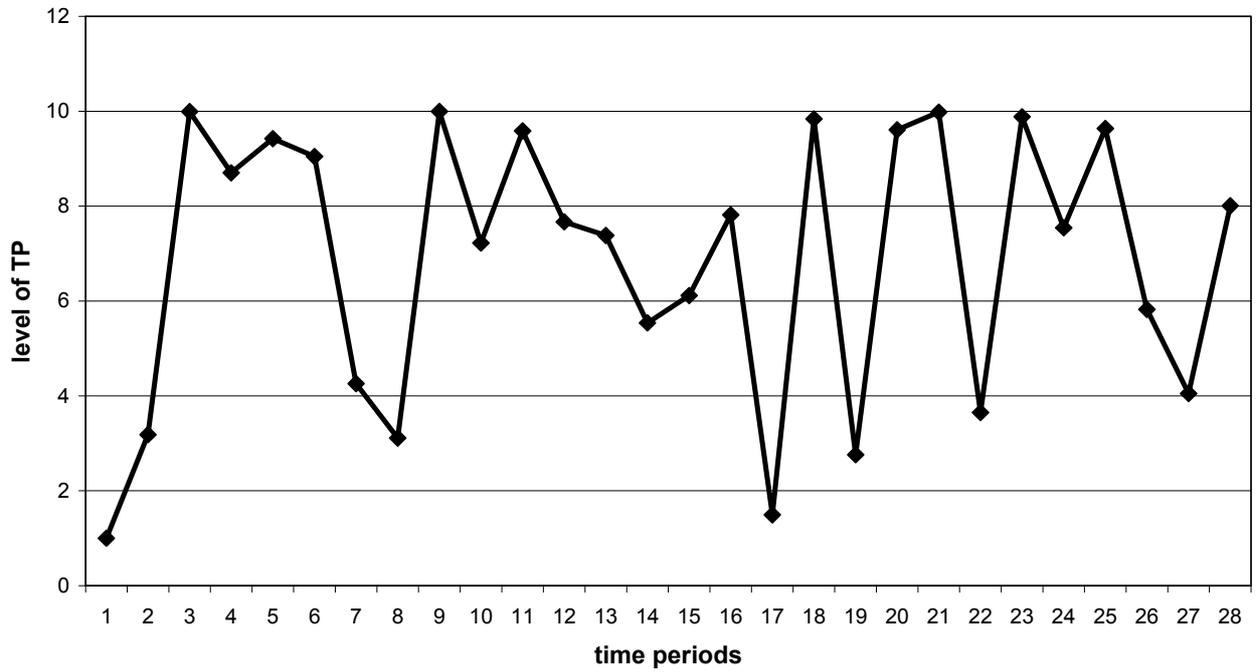

**4 - TP relative to graph 2**

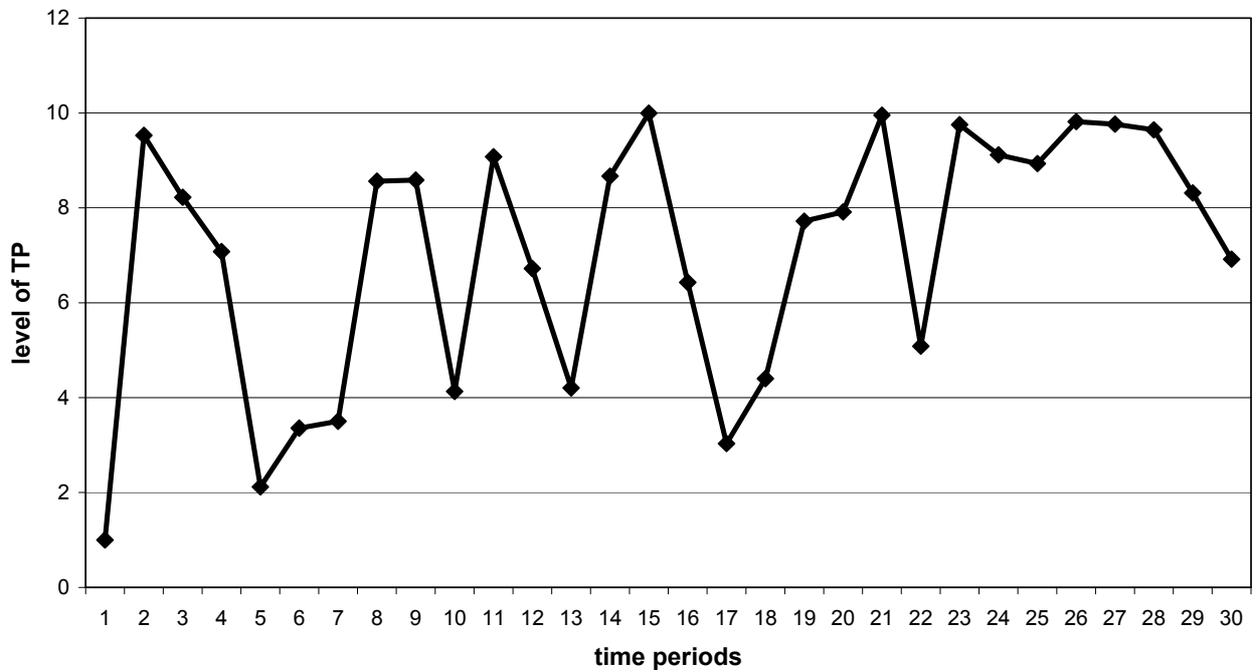

**5 - TP relative to graph 3**

## 3. The next stage

The computer revolution is far from been a novelty. If we set the beginning of it from the invention of the personal computer, it's already two decades old. During this time span the information technology moved from research laboratories to production lines and people's houses. Using a Darwinian frame of interpretation, it can be said that such an evolutionary process started from a "mutation" – the invention of the talented amateurs of the "Homebrew Computer Club" – which proved itself to be the fittest and therefore it held the power to transmute every aspect of society: not only the technical and the economical ones, but also those involving the artistic side of live. If it's obvious that the financial "bubble" of the Clintonian age – with its millenaristic expectations and the lust for easy money – was heavily influenced by the advent of the new technology, it should be noticed that without it artists like William Gibson would never have come into existence. The computer revolution acted on our minds as it did on the world. It modifies our psyche.
Since thirty years are already gone, we could be induced to think that the information technology has provided us with most of what it could be provided. And we may be tempted to shift our attention on the new that is dawning: genetic engineering and nanotechnologies. Wrong idea! It's certainly true that in the near future those innovative technologies will play a central role in the evolution of our society, but the computer revolution is far from been over. There will be new inventions that will supply us with new tools and, what is more relevant, we will experience the saturation of the characteristics of the information technology which are already in existence but have not yet expressed their full potentials. Let's enumerate some of these aspects.
1. The storage – i.e. the ability to preserve information in a digital form. In the last years the cost of storage has been dramatically decreasing, while the capacity and the speed of the devices have been increasing at a very fast speed. In the near future these trends will continue with a growing intensity.
2. The internet. The synapses of the net are spreading all over the world and the speed of data transfer is limited only by the will to implement technologies that are already available.
3. The computing power. While the processing units are still getting more and more powerful, technologies like the one employed in the supercomputers and in the grid computing allow for unlimited computing power.

In order to fully exploit the above sketched hardware structures, we will need advanced software tools which should be able to operate with a high degree of autonomy: in other word, we need to develop that experimental branch of learning that falls under the category of artificial intelligence. Such an aspect is really the bottleneck of the process: obviously developing the hardware is easier than creating the software. Still, even if at a slower pace, the forth corner of the building will be created and a global intelligent system will eventually come into existence.
Such entity will be *ubiquitous*, because it will be able to move instantly on the internet.

It will be *omniscient* by having access to every database in the world.
And since every significant apparatus will depend eventually on control lines traveling on the internet, it will be *omnipotent*.
It goes without saying it, it will be *very intelligent*.
An artificial god will be born.
Will it be like the one in Gibson's "Neuromancer"?

---===<ooo0O0ooo>===---

Predicting the future is a difficult task… Whether the new artificial god will be benevolent or malignant, it will be probably a matter of opinion. But one point can be affirmed: there will be available the flux of information which is necessary to build a full-scale realistic simulation of the kind that as just been sketched. In fact, in order to turn the "proof of concept" into an operative forecasting tool, two components are needed. Firstly, we should pursue a detailed reconstruction of the inner nature of each agent that appears in the simulation. Every entity should be treated as if it were a living being endorsed with self-consciousness. We should model the inner structures of our agents with the same care that a Jungian psychoanalyst employs to decipher the archetypal structures of his patient's subconscious. And we should define the complex network that allows each object to communicate with the rest of the virtual world, setting entrances which let the outside in and the doorways that convey the inside out.
Secondly, we need every available piece of information about the phenomena that are going to be simulated, in order to adjust the virtual agents to their counterparts existing in the real world.
Our artificial god will provide the second element, while the first one is up to us…

# References to the author's writings

PEARCE D., TUCCI M., *A General Net Structure for Theoretical Economics*, in: STEGMÜLLER W., BALZER W., SPOHN W. (eds.), *Philosophy of Economics*, Springer - Verlag, Berlin 1982.

TUCCI M., *Intertheory Relations in Classical Economics: Sraffa and Marx*, "Materiale di Discussione del Dipartimento di Economia Pubblica", n. 12, University of Rome "La Sapienza", Roma 1992.

TUCCI M., *La difformità dei tassi di profitto: un'analisi teorica*, in: MARZANO F. (ed.), *Differenziali e rendite nella distribuzione funzionale del reddito*, La Sapienza, Roma 1996.

TUCCI M., *L'equilibrio economico generale: qualche considerazione sull'evoluzione del paradigma a partire da "Theory of Value" di G. Debreu*, "Materiale di Discussione del Dipartimento di Economia Pubblica", n. 30, University of Rome "La Sapienza", Roma 1997.

TUCCI M., *Competizione, evoluzione della struttura tecnologica e difformità del saggio di profitto*, in: GAROFALO G., PEDONE A. (eds.), *Distribuzione, redistribuzione e crescita. Gli effetti delle diseguaglianze distributive*, FrancoAngeli, Milano 2000.

TUCCI M., BARCHETTI L., *L'IPE: un indice computabile della pressione evolutiva operante sui mercati*, "Working Paper del Dipartimento di Economia Pubblica", n. 43, University of Rome "La Sapienza", Roma 2001.

TUCCI M., *Evolution and Gravitation: a Computer Simulation of a Non-Walrasian Equilibrium Model*, Discussion Paper - September 2002, published on the E-print Archives at arXiv.com (section: Computer Science, registration number: cs.CY/0209017). The text is available at the following address: http://arxiv.org/abs/cs.CY/0209017

TUCCI M., *Oligopolistic Competition in an Evolutionary Environment: a Computer Simulation*, Discussion Paper - January 2005, published on the E-print Archives at arXiv.com (section: Computer Science, registration number: cs.CY/0501037). The text is available at the following address: http://arxiv.org/abs/cs.CY/0501037

Note. The texts of all the papers are available for download at the following addresses:
http://dep.eco.uniroma1.it/~tucci/
http://xoomer.virgilio.it/michele_tucci/

# Bibliography

DARWIN C., *The Origin of Species*.

DEBREU G., *Theory of Value*.

DE FINETTI B., the complete works, especially what concerns subjective probability.

GIBSON W., *New Rose Hotel, Neuromancer*.

HERACLITUS, fragments.

Hicks J.R., the complete works, especially what concerns "hicksian weeks".

Jung C.G., the works about archetypes.

Levy S., *Hackers. Heroes of the computer revolution*.

Keynes J.M., *The General Theory of Employment, Interest and Money*, especially *Chapter 12 – The State of Long-Term Expectation*.

Parmenides, fragments.

Poincaré J.H., the methodological works.

Prigogine I., the methodological works.

Simon H.A., the complete works, especially what concerns the simulation principles.

Schumpeter J.A., the complete works, especially what concerns the evolution of productive structures.

Sraffa P., *Production of Commodities by Means of Commodities*.

Stroustrup B., *The C++ Programming Language*.

Thom R., the works about catastrophe theory.

Walras L., the works about general equilibrium theory.

Wittgenstein L., *Tractatus Logico-Philosophicus*.